\documentclass[12pt]{iopart}

\usepackage{iopams, graphicx}

\begin{document}

\title{Numerical study of curvature perturbations in a brane-world inflation
at high-energies}

\author{Takashi Hiramatsu$^1$ and Kazuya Koyama$^2$}

\address{$^1$Department of Physics , University of Tokyo, Tokyo 113-0033, Japan}
\address{$^2$Institute of Cosmology and Gravitation, University of
 Portsmouth, Portsmouth PO1 2EG, United Kingdom}
\ead{hiramatsu\_at\_utap.phys.s.u-tokyo.ac.jp, Kazuya.Koyama\_at\_port.ac.uk}

\begin{abstract}
We study the evolution of scalar curvature perturbations 
in a brane-world inflation model in a 5D Anti-de Sitter spacetime. The inflaton
perturbations are confined to a 4D brane but they are coupled to the 5D
bulk metric perturbations. We numerically solve full coupled equations
for the inflaton perturbations and the 5D metric perturbations
using Hawkins-Lidsey inflationary model. 
At an initial time, we assume that the bulk is unperturbed.
We find that the inflaton perturbations at high energies
are strongly coupled to the bulk metric perturbations even on subhorizon
scales, leading to the suppression of the amplitude of the comoving 
curvature perturbations at a horizon crossing. 
This indicates that 
the linear perturbations of the inflaton field does not obey the usual 4D
Klein-Gordon equation
due to the coupling to 5D gravitational field on
small scales and it is required to quantise the coupled brane-bulk
system in a consistent way in order to calculate the spectrum of the 
scalar perturbations in a brane-world inflation.
\end{abstract}

\submitto{Journal of Cosmology and Astroparticle Physics}

%===========================================================================%
%===========================================================================%
\section{Introduction}
\label{sec:introduction}
%===========================================================================%
%===========================================================================%
Our view of how our Universe may be described within a higher-dimensional 
spacetime has been revolutionised via brane-world
models \cite{soliton} (see, c.g., \cite{Rev_Langlois, Review, Rev_Brax} for review).
A simple way to study the effects of the higher-dimensional gravity on
inflation models is to consider a 4D inflaton field confined to the
brane in the context of the Randall-Sundrum model \cite{MWBH}. Work to
date has concentrated mainly on deriving corrections to the standard 4D
result by assuming that 5D metric perturbations generated by the 
inflaton perturbations can be neglected.
Then the modification of spectrum is coming from the
modified Friedmann equation. Although a lot of interesting results have
been obtained in this direction \cite{modified, obs}, it remains to see
the consistency of neglecting the bulk metric perturbations. The
analysis of the evolution of gravitational waves in this model has
revealed that the creation of massive graviton modes in the bulk is as
important as the modification of the Friedmann equation \cite{GW}. In
fact, these two higher-dimensional effects cancel out to give a nearly
identical spectrum with the conventional 4D models for the stochastic
background of gravitational waves in the radiation dominated universe
\cite{HT}. This highlights the importance of taking into account the
higher-dimensional effects consistently.

The study of the effects of bulk metric perturbations has been initiated 
in Ref.~\cite{KLMW, KMW} (see \cite{alt} for a different approach). 
Using the approximation that the geometry of the brane is approximated 
by de Sitter spacetime for a slow-roll inflation, the bulk metric perturbations
sourced by inflaton fluctuations on the brane were calculated. It turns out
that the inflaton fluctuations excite an infinite ladder of massive modes of 
bulk metric perturbations. Then Ref.~\cite{KMW} tried to estimate the
effects of the backreaction of bulk metric perturbations. Assuming that
the inflaton field behaves in the same ways as 4D models on sufficiently
small scales at the 0-th order of slow-roll parameters, the corrections
to the evolution of inflaton perturbations at the first order of
slow-roll parameters were calculated. Unlike in 4D models where the
slow-roll corrections can be completely ignored on sub-horizon scales,
the slow-roll corrections play a role even on sub-horizon scales,
yielding the change of the amplitude at a horizon crossing. This result
may not be so surprising as the bulk metric perturbations are strongly
coupled to inflaton perturbations on small scales where gravity becomes
5D. However, this indicates that we should carefully check the validity
of using the 4D formula just by modifying the background Friedmann equations.

This paper focuses on the classical evolution of inflaton perturbations 
coupled to bulk metric perturbations. We do not assume de Sitter geometry 
on the brane and we take into account the backreaction 
of inflaton dynamics consistently. This entails a numerical analysis to 
solve the coupled system directly. We investigate whether the inflaton 
perturbations behave as free massless fields on small scales.

The structure of this paper is as follows. In section II, we set up 
basic equations to describe the evolution of inflaton perturbations
coupled to metric perturbations. We use the Hawkins-Lidsey model of the 
inflation on the brane where the background solution is derived analytically 
\cite{HL}. The Gaussian-normal coordinate is used to solve the perturbations.
Section III is devoted to the numerical scheme. In section IV,
we explain initial conditions and boundary conditions. Then 
we study the evolution of the comoving curvature perturbations, 
which are directly related to observables.

%===========================================================================%
%===========================================================================%
\section{Basic equations}
\label{sec:basic}
%===========================================================================%
%===========================================================================%

%//////////////////////////////////////////////////////%
%
\subsection{Background spacetime}
\label{subsec:background}
%
%//////////////////////////////////////////////////////%

We consider the Randall-Sundrum single-brane model \cite{RS}. In this model, a
3D spatial hypersurface is embedded in a 5D anti-de
Sitter (AdS$_5$) space-time with a negative cosmological constant
$\Lambda_5$. The 5D Einstein equations in the bulk are given by
%%%%%%%%%%%%%%%%%%%%%%%%%%%%%%%%%%%%%%%%%%%%%%%%%%%%%%%%%%%%%%%%%%%%%%%%%
%
\begin{equation}
 {}^{(5)}G_{AB}+\Lambda_5{}^{(5)}g_{AB}=0.
 \label{eq:5D_Einstein}
\end{equation}
%
%%%%%%%%%%%%%%%%%%%%%%%%%%%%%%%%%%%%%%%%%%%%%%%%%%%%%%%%%%%%%%%%%%%%%%%%%
The cosmological constant is related with the curvature scale of the AdS$_5$
space-time $\mu$ as $\Lambda_5=-6\mu^2$. The embedded brane has a tension
$\lambda$ given by $\kappa_5^2\lambda=6\mu$. Here
$\kappa_5$ denotes the 5D gravitational constant which can be
rewritten using the 5D Planck mass $M_5$ as
$\kappa_5^2=8\pi/M_5^3$. On the other hand, we use $\kappa_4$ as
4D one defined as $\kappa_4^2=8\pi/M_{\rm pl}^2$ which is
related with $\kappa_5$ as $\kappa_4^2=\mu\kappa_5^2$, where $M_{\rm pl}$
is the 4D Planck mass. 
The induced metric on the brane is defined as 
%%%%%%%%%%%%%%%%%%%%%%%%%%%%%%%%%%%%%%%%%%%%%%%%%%%%%%%%%%%%%%%%%%%%%%%%%
%
\begin{equation}
  g_{AB} ={}^{(5)}g_{AB}-n_An_B, 
  \label{eq:induced_metric}
\end{equation}
%
%%%%%%%%%%%%%%%%%%%%%%%%%%%%%%%%%%%%%%%%%%%%%%%%%%%%%%%%%%%%%%%%%%%%%%%%%
where $n^A$ denotes the unit normal vector to the brane.
The energy-momentum tensor $T_{\mu\nu}$ on the brane and a brane
tension $-\lambda g_{\mu\nu}$ produce the jump in the extrinsic
curvature at the brane;
%%%%%%%%%%%%%%%%%%%%%%%%%%%%%%%%%%%%%%%%%%%%%%%%%%%%%%%%%%%%%%%%%%%%%%%%%
%
\begin{equation}
  K^\mu{}_\nu{}^+ - K^\mu{}_\nu{}^- = 2K^\mu{}_\nu{}^+ =
   -\kappa_5^2\left[T^\mu{}_\nu -
	       \frac{1}{3}g^\mu{}_\nu(T-\lambda)\right],
  \label{eq:K_jump}
\end{equation}
%
%%%%%%%%%%%%%%%%%%%%%%%%%%%%%%%%%%%%%%%%%%%%%%%%%%%%%%%%%%%%%%%%%%%%%%%%%
where we imposed the $Z_2$ symmetry on the brane and the extrinsic
curvature is defined by $K_{\mu\nu} = g_\mu{}^Cg_\nu{}^D{}^{(5)}\nabla_Cn_D$.
According to Ref.~\cite{SMS}, the effective Einstein equation for the
induced metric on the brane is obtained as 
%%%%%%%%%%%%%%%%%%%%%%%%%%%%%%%%%%%%%%%%%%%%%%%%%%%%%%%%%%%%%%%%%%%%%%%%%
%
\begin{equation}
  {}^{(4)}G_{\mu\nu}=\kappa_4^2T_{\mu\nu}+\kappa_5^4\Pi_{\mu\nu}-E_{\mu\nu},
  \label{eq:effective_Einstein}
\end{equation}
%
%%%%%%%%%%%%%%%%%%%%%%%%%%%%%%%%%%%%%%%%%%%%%%%%%%%%%%%%%%%%%%%%%%%%%%%%%
where $\Pi_{\mu\nu}$ is the quadratic terms of the energy-momentum
tensor,
%%%%%%%%%%%%%%%%%%%%%%%%%%%%%%%%%%%%%%%%%%%%%%%%%%%%%%%%%%%%%%%%%%%%%%%%%
%
\begin{equation}
 \Pi_{\mu\nu} = -\frac{1}{4}T_{\mu\alpha}T_\nu{}^\alpha +
  \frac{1}{12}TT_{\mu\nu} +
  \frac{1}{8}g_{\mu\nu}T_{\alpha\beta}T^{\alpha\beta} -
  \frac{1}{24}g_{\mu\nu}T^2,
  \label{eq:Pi}
\end{equation}
%
%%%%%%%%%%%%%%%%%%%%%%%%%%%%%%%%%%%%%%%%%%%%%%%%%%%%%%%%%%%%%%%%%%%%%%%%%
and $E_{\mu\nu}$ is the electric part of 5D Weyl tensor
$C_{ABCD}$,
%%%%%%%%%%%%%%%%%%%%%%%%%%%%%%%%%%%%%%%%%%%%%%%%%%%%%%%%%%%%%%%%%%%%%%%%%
%
\begin{equation}
 E_{\mu\nu} = C^{E}{}_{AFB}n_En^Fg_{\mu}{}^Ag_{\nu}{}^B.
  \label{eq:projected_Weyl}
\end{equation}
%
%%%%%%%%%%%%%%%%%%%%%%%%%%%%%%%%%%%%%%%%%%%%%%%%%%%%%%%%%%%%%%%%%%%%%%%%%

We assume that the matter content confined to the brane is described by
a perfect fluid whose energy-momentum tensor is given by 
%%%%%%%%%%%%%%%%%%%%%%%%%%%%%%%%%%%%%%%%%%%%%%%%%%%%%%%%%%%%%%%%%%%%%%%%%
%
\begin{equation}
 T^{\mu}{}_{\nu}= {\rm diag}(-\rho,p,p,p),
   \label{eq:energy_momentum}
\end{equation}
%
%%%%%%%%%%%%%%%%%%%%%%%%%%%%%%%%%%%%%%%%%%%%%%%%%%%%%%%%%%%%%%%%%%%%%%%%%
where $\rho$ and $p$ denote the energy density and the pressure of matter on
the brane, respectively.
In this paper, we use the Gaussian-Normal (GN) coordinate. The metric
is given by \cite{BDEL}
%%%%%%%%%%%%%%%%%%%%%%%%%%%%%%%%%%%%%%%%%%%%%%%%%%%%%%%%%%%%%%%%%%%%%%%%%
%
\begin{equation}
 ds^2 =  - n^2(y,t)dt^2 + a^2(y,t)\delta_{ij}dx^i dx^j + b^2(y,t)dy^2.
 \label{eq:background_metric}
\end{equation}
%
%%%%%%%%%%%%%%%%%%%%%%%%%%%%%%%%%%%%%%%%%%%%%%%%%%%%%%%%%%%%%%%%%%%%%%%%%
Here the warp function $a(y,t)$ and the lapse function $n(y,t)$ are
obtained in \cite{BDEL} as
%%%%%%%%%%%%%%%%%%%%%%%%%%%%%%%%%%%%%%%%%%%%%%%%%%%%%%%%%%%%%%%%%%%%%%%%%
%
\begin{eqnarray}
 a(y,t) &= a_0(t)\left\{\cosh\mu y -
 \left(1+\frac{\kappa_5^2\rho}{6\mu}\right)\sinh\mu y\right\},\\
 n(y,t) &= \cosh\mu y -
 \left\{1-\frac{\kappa_5^2}{6\mu}(2 \rho+3p)\right\}\sinh\mu y,
 \label{eq:lapse_warp}
\end{eqnarray}
%
%%%%%%%%%%%%%%%%%%%%%%%%%%%%%%%%%%%%%%%%%%%%%%%%%%%%%%%%%%%%%%%%%%%%%%%%%
where $a_0$ is the scale factor on the brane, and we can set
$b(y,t)=1$ without a loss of generality. The scale factor is determined by the
Friedmann equation and the energy conservation law as
%%%%%%%%%%%%%%%%%%%%%%%%%%%%%%%%%%%%%%%%%%%%%%%%%%%%%%%%%%%%%%%%%%%%%%%%%
%
\begin{eqnarray}
  H^2 &= \frac{\kappa_4^2}{3}\rho\left(1+\frac{\rho}{2\lambda}\right),
 \label{eq:Friedmann_equation} \\
  \dot{\rho} &= -3(p+\rho)H.
 \label{eq:conservation_law}
\end{eqnarray}
%
%%%%%%%%%%%%%%%%%%%%%%%%%%%%%%%%%%%%%%%%%%%%%%%%%%%%%%%%%%%%%%%%%%%%%%%%%
In this coordinate, the brane is located at $y=y_{\rm b}=0$. 
Note that, in this coordinate, there is a coordinate singularity
at the finite distance from the brane,
%%%%%%%%%%%%%%%%%%%%%%%%%%%%%%%%%%%%%%%%%%%%%%%%%%%%%%%%%%%%%%%%%%%%%%%%%
%
\begin{equation}
y_{\rm c} =\frac{1}{\mu}\coth^{-1}\left(\frac{H}{\mu}\right),
 \label{eq:singularity}
\end{equation}
%
%%%%%%%%%%%%%%%%%%%%%%%%%%%%%%%%%%%%%%%%%%%%%%%%%%%%%%%%%%%%%%%%%%%%%%%%%
where the warp function vanishes $a(y_{\rm c},t)=0$. 
The location of the coordinate singularity $y_{\rm c}$ agrees with the 
Cauchy horizon of the AdS$_5$ spacetime \cite{coordinate}.  
We will come back to this issue later.

In this paper, we consider a scalar field on the brane. 
The scalar field satisfies the equation of motion 
%%%%%%%%%%%%%%%%%%%%%%%%%%%%%%%%%%%%%%%%%%%%%%%%%%%%%%%%%%%%%%%%%%%%%%%%%
%
\begin{equation}
  \ddot{\phi} + 3H\dot{\phi} + V'(\phi) = 0,
     \label{eq:scalar_equation}
\end{equation}
%
%%%%%%%%%%%%%%%%%%%%%%%%%%%%%%%%%%%%%%%%%%%%%%%%%%%%%%%%%%%%%%%%%%%%%%%%%
where $V(\phi)$ denotes the potential of the scalar field.
We use a model proposed by Hawkins-Lidsey as an inflaton model on the brane
to avoid numerical integrations of the background equations \cite{HL}.
The potential for inflaton is given by 
%%%%%%%%%%%%%%%%%%%%%%%%%%%%%%%%%%%%%%%%%%%%%%%%%%%%%%%%%%%%%%%%%%%%%%%%%
%
\begin{equation}
  V(\phi)= \frac{\lambda}{3} (6-C^2) {\rm cosech}^2 
    \left(\frac{\sqrt{2 \pi} C}{M_{\rm pl}} \phi \right),
\end{equation}
%
%%%%%%%%%%%%%%%%%%%%%%%%%%%%%%%%%%%%%%%%%%%%%%%%%%%%%%%%%%%%%%%%%%%%%%%%%
where $C$ is constant.
An advantage of this potential is that an exact solution for the scale
factor is known
%%%%%%%%%%%%%%%%%%%%%%%%%%%%%%%%%%%%%%%%%%%%%%%%%%%%%%%%%%%%%%%%%%%%%%%%%
%
\begin{equation}
  a_0(t)= \left[(C^2 \mu t+1)^2-1\right]^{1/C^2}.
\end{equation}
%
%%%%%%%%%%%%%%%%%%%%%%%%%%%%%%%%%%%%%%%%%%%%%%%%%%%%%%%%%%%%%%%%%%%%%%%%%
The energy density and pressure of the scalar field can be
calculated as
%%%%%%%%%%%%%%%%%%%%%%%%%%%%%%%%%%%%%%%%%%%%%%%%%%%%%%%%%%%%%%%%%%%%%%%%%
%
\begin{eqnarray}
  \rho &= \frac{2 \lambda}{(C^2 \mu t +1)^2 -1}, \\
     p &= \frac{2 \lambda}{3}(C^2-3)\frac{1}{(C^2 \mu t+1)^2 -1}.
\end{eqnarray}
%
%%%%%%%%%%%%%%%%%%%%%%%%%%%%%%%%%%%%%%%%%%%%%%%%%%%%%%%%%%%%%%%%%%%%%%%%%
Thus the effective equation of state $w$ becomes
%%%%%%%%%%%%%%%%%%%%%%%%%%%%%%%%%%%%%%%%%%%%%%%%%%%%%%%%%%%%%%%%%%%%%%%%%
%
\begin{equation}
  w = \frac{p}{\rho} = -1 +\frac{C^2}{3}.
\end{equation}
%
%%%%%%%%%%%%%%%%%%%%%%%%%%%%%%%%%%%%%%%%%%%%%%%%%%%%%%%%%%%%%%%%%%%%%%%%%
Then for small $C^2$, a slow-roll inflation is realized, which is a
generalisation of the power-law inflation. 

%//////////////////////////////////////////////////////%
%
\subsection{Scalar perturbations}
\label{subsec:perturbation}
%
%//////////////////////////////////////////////////////%

%-------------------------------------%
%
\subsubsection{Equations in the bulk}
\label{subsec:equations_bulk}
%
%-------------------------------------%
Now let us consider the scalar perturbations. There are several ways to
calculate the scalar perturbations in a AdS$_5$ spacetime. Here we describe an
approach based on a master variable \cite{Mukohyama, Kodama}. Using the
generalized 5D longitudinal gauge, the perturbed spacetime is given by 
%%%%%%%%%%%%%%%%%%%%%%%%%%%%%%%%%%%%%%%%%%%%%%%%%%%%%%%%%%%%%%%%%%%%%%%%%
%
\begin{eqnarray}
  ds^2 &=& -n(y,t)^2(1+2 A) dt^2+a(y,t)^2(1+2 {\cal R})\delta_{ij}
  dx^i dx^j \nonumber \\
        && +(1+2 A_{yy})dy^2+n(y,t)A_{y}dydt.
  \label{eq:perturbed_metric}
\end{eqnarray}
%
%%%%%%%%%%%%%%%%%%%%%%%%%%%%%%%%%%%%%%%%%%%%%%%%%%%%%%%%%%%%%%%%%%%%%%%%%
It was shown that the solution for metric perturbations can be 
derived from a master variable $\Omega$ as % \cite{Mu, KIS, p16, p21}
%%%%%%%%%%%%%%%%%%%%%%%%%%%%%%%%%%%%%%%%%%%%%%%%%%%%%%%%%%%%%%%%%%%%%%%%%
%
\begin{eqnarray}
 A &=  -\frac{1}{6a}
    \left\{ \left( 2\Omega'' - \frac{n'}{n} \Omega' \right)
     + \frac{1}{n^2} \left( \ddot\Omega - \frac{\dot n}{n} \dot\Omega \right)
     - \mu^2 \Omega
    \right\}\, ,  \label{eq:metric_A}
    \\
 A_y &=
     \frac{1}{na}\left( \dot\Omega'-\frac{n'}{n}\dot\Omega \right)\, ,  
     \label{eq:metric_Ay} \\
 A_{yy} &= \frac{1}{6a} \left\{
     \left( \Omega''- 2\frac{n'}{n} \Omega' \right)
      + \frac{2}{n^2} \left( \ddot\Omega - \frac{\dot{n}}{n} \dot\Omega\right)
      + \mu^2 \Omega \right\} \,, \label{eq:metric_Ayy} \\
{\cal R} &=
    \frac{1}{6a}\left\{ \left( \Omega''+ \frac{n'}{n}\Omega' \right)
      +\frac{1}{n^2}\left(-\ddot\Omega + \frac{\dot n}{n} \dot\Omega \right)
      - 2\mu^2 \Omega\right\} \,, \label{eq:metric_R}
\end{eqnarray}
%
%%%%%%%%%%%%%%%%%%%%%%%%%%%%%%%%%%%%%%%%%%%%%%%%%%%%%%%%%%%%%%%%%%%%%%%%%
as long as the master variable $\Omega$ in the bulk satisfies a wave
equation given by
%%%%%%%%%%%%%%%%%%%%%%%%%%%%%%%%%%%%%%%%%%%%%%%%%%%%%%%%%%%%%%%%%%%%%%%%%
%
\begin{equation}
 - \left( \frac{1}{na^3} \dot\Omega \right)^\cdot
 + \left( \frac{n}{a^3} \Omega' \right)^\prime
 + \left( \mu^2 + \frac{\nabla^2}{a^2} \right)
 \frac{n}{a^3} \Omega = 0 \,,
 \label{eq:master_equation}
\end{equation}
%
%%%%%%%%%%%%%%%%%%%%%%%%%%%%%%%%%%%%%%%%%%%%%%%%%%%%%%%%%%%%%%%%%%%%%%%%%
where a prime and a dot denote the derivatives with respect to $y$ and
$t$, respectively.

%-------------------------------------%
%
\subsubsection{Boundary conditions}
\label{subsec:equations_boundary}
%
%-------------------------------------%
In this section, we derive the evolution equation for inflaton perturbations
coupled to the master variable. See Refs.~\cite{KLMW, gaugeLMSW,
gaugeBMW, gaugeD, gaugeK} and references therein for earlier works.
We begin with the effective Einstein equations on the brane \cite{SMS}.
The perturbed effective Einstein equations are given by 
%%%%%%%%%%%%%%%%%%%%%%%%%%%%%%%%%%%%%%%%%%%%%%%%%%%%%%%%%%%%%%%%%%%%%%%%%
%
\begin{equation}
  {}^{(4)}\delta G_{\mu\nu} = \kappa_4^2\delta T_{\mu\nu}
     + \kappa_5^4\delta \Pi_{\mu\nu}-\delta E_{\mu\nu}.
  \label{eq:pert_effective_Einstein}
\end{equation}
%
%%%%%%%%%%%%%%%%%%%%%%%%%%%%%%%%%%%%%%%%%%%%%%%%%%%%%%%%%%%%%%%%%%%%%%%%%
The perturbed Einstein tensor is given by \cite{gaugeBMW}
%%%%%%%%%%%%%%%%%%%%%%%%%%%%%%%%%%%%%%%%%%%%%%%%%%%%%%%%%%%%%%%%%%%%%%%%%
%
\begin{eqnarray}
  {}^{(4)}\delta G^0{}_0 &= 6H\left( -\dot{\cal R}_{\rm b}+HA_{\rm b} \right)
                  + \frac{2}{a_0^2}\nabla^2{\cal R}_{\rm b}\;, \\
  {}^{(4)}\delta G^0{}_i &= -2(HA_{\rm b}-\dot{\cal R}_{\rm b})_{,i}\;, \\
  {}^{(4)}\delta G^i{}_j &= \delta^i{}_j{}^{(4)}\delta G_T 
      + \left(\delta G_{TF}{}^{,i}{}_{,j}
              -\frac{1}{3}\delta^i{}_j\delta G_{TF}{}^{,k}{}_{,k} \right)\;,
  \label{eq:perturbed_G}
\end{eqnarray}
%
%%%%%%%%%%%%%%%%%%%%%%%%%%%%%%%%%%%%%%%%%%%%%%%%%%%%%%%%%%%%%%%%%%%%%%%%%
where
%%%%%%%%%%%%%%%%%%%%%%%%%%%%%%%%%%%%%%%%%%%%%%%%%%%%%%%%%%%%%%%%%%%%%%%%%
%
\begin{eqnarray}
\fl  {}^{(4)}\delta G_{T}  &= 
      2\left\{ (3H^2+2\dot{H})A_{\rm b} + H\dot{A}_{\rm b} 
        - \ddot{\cal R}_{\rm b} - 3H\dot{\cal R}_{\rm b} 
        + \frac{1}{3a_0^2}\nabla^2({\cal R}_{\rm b}+A_{\rm b}) \right\},  \\
\fl  {}^{(4)}\delta G_{TF} &= -\frac{1}{a_0^2}({\cal R}_{\rm b}+A_{\rm b}).
  \label{eq:perturbed_GT_GTF}
\end{eqnarray}
%
%%%%%%%%%%%%%%%%%%%%%%%%%%%%%%%%%%%%%%%%%%%%%%%%%%%%%%%%%%%%%%%%%%%%%%%%%
The subscript b denotes a quantity evaluated on the brane.
We can parameterize the projected Weyl tensor $\delta E_{\mu\nu}$ 
as an effective fluid;
%%%%%%%%%%%%%%%%%%%%%%%%%%%%%%%%%%%%%%%%%%%%%%%%%%%%%%%%%%%%%%%%%%%%%%%%%
%
\begin{eqnarray}
 \delta E^{0}{}_{0} &= \kappa_4^2\delta \rho_{\cal E}~,
   \label{eq:perturbed_E00}  \\
 \delta E^{0}{}_{i} &= -\kappa_4^2\delta q_{{\cal E},i}~,
   \label{eq:perturbed_E0i}  \\
 \delta E^{i}{}_{j} &= 
  -\kappa_4^2(\delta p_{\cal E}\delta^i{}_{j}+\delta\pi_{\cal E}{}^i{}_{j})~,
   \label{eq:perturbed_Eij}
\end{eqnarray}
%
%%%%%%%%%%%%%%%%%%%%%%%%%%%%%%%%%%%%%%%%%%%%%%%%%%%%%%%%%%%%%%%%%%%%%%%%%
where
%%%%%%%%%%%%%%%%%%%%%%%%%%%%%%%%%%%%%%%%%%%%%%%%%%%%%%%%%%%%%%%%%%%%%%%%%
%
\begin{equation}
 \delta\pi_{\cal E}{}^i{}_{j} = \delta\pi_{\cal E}{}^{,i}{}_{,j}
    -\frac{1}{3}\delta^i{}_j\delta\pi_{\cal E}{}^{,k}{}_{,k}~.
 \label{eq:Weyl_pi}
\end{equation}
%
%%%%%%%%%%%%%%%%%%%%%%%%%%%%%%%%%%%%%%%%%%%%%%%%%%%%%%%%%%%%%%%%%%%%%%%%%
The Weyl fluid satisfies the radiation-like equation-of-state
$\delta p_{\cal E}=\delta \rho_{\cal E}/3$. 
The perturbed energy-momentum tensor $\delta T_{\mu\nu}$ is defined as
%%%%%%%%%%%%%%%%%%%%%%%%%%%%%%%%%%%%%%%%%%%%%%%%%%%%%%%%%%%%%%%%%%%%%%%%%
%
\begin{eqnarray}
 \delta T^{0}{}_{0} &= -\delta\rho~,
   \label{eq:perturbed_T00}  \\
 \delta T^{0}{}_{i} &= \delta q_{,i}~,
   \label{eq:perturbed_T0i}  \\
 \delta T^{i}{}_{j} &= \delta p \delta^i{}_j~.
   \label{eq:perturbed_Tij} 
\end{eqnarray}
%
%%%%%%%%%%%%%%%%%%%%%%%%%%%%%%%%%%%%%%%%%%%%%%%%%%%%%%%%%%%%%%%%%%%%%%%%%
Throughout this paper, we focus on the perturbations of the
scalar field. Thus the anisotropic stress $\delta\pi^i{}_j$ on the brane
vanishes.
The quadratic energy-momentum tensor $\delta\Pi_{\mu\nu}$ is calculated as 
\cite{gaugeBMW}
%%%%%%%%%%%%%%%%%%%%%%%%%%%%%%%%%%%%%%%%%%%%%%%%%%%%%%%%%%%%%%%%%%%%%%%%%
%
\begin{eqnarray}
 \delta \Pi^{0}{}_{0} &= -\frac{1}{6}\rho\delta\rho~,
   \label{eq:perturbed_Pi00}  \\
 \delta \Pi^{0}{}_{i} &= \frac{1}{6}\rho\delta q_{,i}~,
   \label{eq:perturbed_Pi0i}  \\
 \delta \Pi^{i}{}_{j} &= 
   \frac{1}{6}\left\{(\rho+p)\delta\rho+\rho\delta p\right\}\delta^{i}{}_{j}~.
   \label{eq:perturbed_Piij} 
\end{eqnarray}
%
%%%%%%%%%%%%%%%%%%%%%%%%%%%%%%%%%%%%%%%%%%%%%%%%%%%%%%%%%%%%%%%%%%%%%%%%%
All perturbative quantities can be expanded by scalar harmonic 
functions $e^{i\mathbf{k}\cdot\mathbf{x}}$, which enables us to replace
the operator 
$\nabla^2$ by $-k^2$. Then, from the effective Einstein equations
(\ref{eq:pert_effective_Einstein}), we can derive the evolution equations for
the perturbative quantities as \cite{gaugeK}
%%%%%%%%%%%%%%%%%%%%%%%%%%%%%%%%%%%%%%%%%%%%%%%%%%%%%%%%%%%%%%%%%%%%%%%%%
%
\begin{eqnarray}
    -\frac{1}{2}\kappa_5^2{\cal H}\delta\rho &=
     3H(\dot{{\cal R}_{\rm b}}-HA_{\rm b}) + \frac{k^2}{a_0^2}{\cal R}_{\rm b}
     -\frac{1}{2}\kappa_4^2 \delta \rho_{{\cal E}},  
     \label{eq:effective_Einstein_1} \\
     -\frac{1}{2}\kappa_5^2{\cal H}\delta p &= 
     -\ddot{\cal R}_{\rm b} - H\left(3
     -\frac{\dot{H}}{{\cal H}^2}\right)\dot{{\cal R}}_{\rm b}
     +H\dot{A}_{\rm b}+\left(2\dot{H} +3H^2
     -\frac{H^2 \dot{H}}{{\cal H}^2}\right)A_{\rm b} \nonumber \\
  &- \frac{1}{3}\frac{k^2}{a_0^2}A_{\rm b} 
     - \frac{1}{3}\left(1-\frac{\dot{H}}{{\cal H}^2}\right)
       \frac{k^2}{a_0^2}{\cal R}_{\rm b}
     - \frac{1}{6}\left(1+\frac{\dot{H}}{{\cal H}^2}\right)\kappa_4^2
       \delta \rho_{{\cal E}},
     \label{eq:effective_Einstein_2} \\
    -\frac{1}{2}\kappa_5^2{\cal H}\delta q &= 
     \dot{{\cal R}}_{\rm b}-HA_{\rm b}-\frac{1}{2}\kappa_4^2 \delta q_{{\cal E}},
     \label{eq:effective_Einstein_3} \\
     0 &= \frac{1}{a_0^2}(A_{\rm b}+{\cal R}_{\rm b})+\kappa_4^2\delta \pi_{\cal E},
     \label{eq:effective_Einstein_4}
\end{eqnarray}
%
%%%%%%%%%%%%%%%%%%%%%%%%%%%%%%%%%%%%%%%%%%%%%%%%%%%%%%%%%%%%%%%%%%%%%%%%%
where we defined
%%%%%%%%%%%%%%%%%%%%%%%%%%%%%%%%%%%%%%%%%%%%%%%%%%%%%%%%%%%%%%%%%%%%%%%%%
%
\begin{equation}
 \mathcal{H}\equiv \left(\frac{a'}{a}\right)_{\rm b} =
   -\mu\sqrt{1+\left(\frac{H}{\mu}\right)^2},
 \label{eq:junction_background_2}
\end{equation}
%
%%%%%%%%%%%%%%%%%%%%%%%%%%%%%%%%%%%%%%%%%%%%%%%%%%%%%%%%%%%%%%%%%%%%%%%%%
and used the relation $p+\rho=2\dot{H}/\kappa_5^2{\cal H}$ derived from
Eqs.~(\ref{eq:Friedmann_equation}) and (\ref{eq:conservation_law}).
Ref.~\cite{gaugeD} showed that the components of the perturbative  Weyl tensor
$\delta\rho_{\cal E},\delta q_{\cal E},\delta\pi_{\cal E}$ can be written
in terms of the master variable $\Omega$ as 
%%%%%%%%%%%%%%%%%%%%%%%%%%%%%%%%%%%%%%%%%%%%%%%%%%%%%%%%%%%%%%%%%%%%%%%%%
%
\begin{eqnarray}
  \kappa_4^2 \delta \rho_{\cal E} &= \left(
     \frac{k^4 \Omega}{3a_0^5}\right)_{\rm b},
     \label{eq:Weyl_1}\\
  \kappa_4^2 \delta q_{{\cal E}} &=- 
     \frac{k^2}{3a_0^3}\left(H\Omega - \dot{\Omega} \right)_{\rm b},
     \label{eq:Weyl_2}\\
  \kappa_4^2 \delta \pi_{{\cal E}} &= 
     \frac{1}{6a_0^3}\left\{3 \ddot{\Omega} - 3H\dot{\Omega}
    +\frac{k^2}{a_0^2}\Omega-3\left(\frac{n'}{n}-\frac{a'}{a}\right)
     \Omega' \right\}_{\rm b}.
     \label{eq:Weyl_3}
\end{eqnarray}
%
%%%%%%%%%%%%%%%%%%%%%%%%%%%%%%%%%%%%%%%%%%%%%%%%%%%%%%%%%%%%%%%%%%%%%%%%%

The perturbed energy-momentum tensor is calculated as 
%%%%%%%%%%%%%%%%%%%%%%%%%%%%%%%%%%%%%%%%%%%%%%%%%%%%%%%%%%%%%%%%%%%%%%%%%
%
\begin{eqnarray}
  \delta \rho &= -\dot{\phi}^2A_{\rm b} + \dot{\phi}\delta\dot{\phi} 
                                + V'(\phi) \delta \phi,
     \label{eq:scalar_energy_momentum_1}\\
  \delta p &= - \dot{\phi}^2A_{\rm b} + \dot{\phi}\delta\dot{\phi} 
                                - V'(\phi) \delta \phi,
     \label{eq:scalar_energy_momentum_2}\\
  \delta q &= - \dot{\phi} \delta \phi.
     \label{eq:scalar_energy_momentum_3}
\end{eqnarray}
%
%%%%%%%%%%%%%%%%%%%%%%%%%%%%%%%%%%%%%%%%%%%%%%%%%%%%%%%%%%%%%%%%%%%%%%%%%
The equation of motion for the scalar field perturbations is derived from the
conservation law $\nabla_\mu\delta T^\mu{}_\nu=0$ as 
%%%%%%%%%%%%%%%%%%%%%%%%%%%%%%%%%%%%%%%%%%%%%%%%%%%%%%%%%%%%%%%%%%%%%%%%%
%
\begin{eqnarray}
  \delta \ddot{\phi} + 3H \delta \dot{\phi} + \frac{k^2}{a_0^2} 
  \delta \phi = - V''(\phi) \delta \phi - 3 \dot{\phi} \dot{{\cal R}}_{\rm b}
                  + \dot{\phi} \dot{A}_{\rm b} - 2 V'(\phi) A_{\rm b}.
  \label{eq:brane_equation}
\end{eqnarray}
%
%%%%%%%%%%%%%%%%%%%%%%%%%%%%%%%%%%%%%%%%%%%%%%%%%%%%%%%%%%%%%%%%%%%%%%%%%
Now we introduce a gauge-invariant variable, the Mukhanov-Sasaki variable,
defined as \cite{Mukhanov}
%%%%%%%%%%%%%%%%%%%%%%%%%%%%%%%%%%%%%%%%%%%%%%%%%%%%%%%%%%%%%%%%%%%%%%%%%
%
\begin{equation}
  Q \equiv \delta \phi - \frac{\dot{\phi}}{H} {\cal R}_{\rm b}.
  \label{eq:Mukhanov-Sasaki}
\end{equation}
%
%%%%%%%%%%%%%%%%%%%%%%%%%%%%%%%%%%%%%%%%%%%%%%%%%%%%%%%%%%%%%%%%%%%%%%%%%
Substituting this definition into Eq.~(\ref{eq:brane_equation}) 
and using the effective Einstein equations, we can derive the
evolution equation for $Q$ as
%%%%%%%%%%%%%%%%%%%%%%%%%%%%%%%%%%%%%%%%%%%%%%%%%%%%%%%%%%%%%%%%%%%%%%%%%
%
\begin{equation} \fl
  \ddot{Q} + 3 H \dot{Q} + \frac{k^2}{a_0^2} Q 
  + \left\{\frac{\ddot{H}}{H} -2 \frac{\dot{H}}{H} \frac{V'(\phi)}{\dot{\phi}}
  -2 \left(\frac{\dot{H}}{H}\right)^2 + V''(\phi) \right\}Q =J(\Omega),
  \label{eq:Mukhanov_Sasaki_equation}
\end{equation}
%
%%%%%%%%%%%%%%%%%%%%%%%%%%%%%%%%%%%%%%%%%%%%%%%%%%%%%%%%%%%%%%%%%%%%%%%%%
where 
%%%%%%%%%%%%%%%%%%%%%%%%%%%%%%%%%%%%%%%%%%%%%%%%%%%%%%%%%%%%%%%%%%%%%%%%%
%
\begin{equation} \fl
J(\Omega)=-\frac{\dot{\phi}}{H} 
\left[   
\left(- \frac{\dot{H}}{H} + \frac{\ddot{H}}{2 \dot{H}} \right) 
\kappa_4^2 \delta q_{{\cal E}} + \frac{1}{3} 
\left(1- \frac{\dot{H}}{2 {\cal H}^2} \right) \kappa_4^2 \delta \rho_{{\cal E}}
+ \frac{1}{3} k^2 \kappa_4^2 \delta \pi_{{\cal E}} 
+ \frac{1}{3} \frac{\dot{H}}{{\cal H}^2} \frac{k^2}{a_0^2} {\cal R}_{\rm b}
\right].
  \label{eq:Mukhanov_Sasaki_source}
\end{equation}
%
%%%%%%%%%%%%%%%%%%%%%%%%%%%%%%%%%%%%%%%%%%%%%%%%%%%%%%%%%%%%%%%%%%%%%%%%%
Here the left-hand side of Eq.~(\ref{eq:Mukhanov_Sasaki_equation}) is 
completely the same as the 4D cosmology 
while $J(\Omega)$ describes the the effect of the bulk metric 
perturbations. Using Eq.~(\ref{eq:master_equation}), 
the curvature perturbations (\ref{eq:metric_R}) evaluated at the brane
${\cal R}_{\rm b}$ can be expressed in terms of $\Omega$ and its
first-order derivatives as
%%%%%%%%%%%%%%%%%%%%%%%%%%%%%%%%%%%%%%%%%%%%%%%%%%%%%%%%%%%%%%%%%%%%%%%%%
%
\begin{equation}
{\cal R}_{\rm b} = \frac{1}{6a_0} 
\left(3{\cal H} \Omega' -3H \dot{\Omega} - 3 \mu^2 \Omega 
+ \frac{k^2}{a_0^2} \Omega \right)_{\rm b}.
  \label{eq:curvature_brane}
\end{equation}
%
%%%%%%%%%%%%%%%%%%%%%%%%%%%%%%%%%%%%%%%%%%%%%%%%%%%%%%%%%%%%%%%%%%%%%%%%%
From the $(0,i)$ component of the effective Einstein equation
(\ref{eq:effective_Einstein_3}), we can relate $Q$ and $\Omega$ as
%%%%%%%%%%%%%%%%%%%%%%%%%%%%%%%%%%%%%%%%%%%%%%%%%%%%%%%%%%%%%%%%%%%%%%%%%
%
\begin{equation}
Q= \frac{1}{\kappa_5^2 \dot{\phi} a_0} 
\left[
  \dot{\Omega}'-{\cal H} \dot{\Omega} - \frac{\dot{H}}{H {\cal H}} 
  \left( {\cal H} \Omega' - \mu^2 \Omega + \frac{k^2}{3 a_0^2} \Omega \right)
\right]_{\rm b},
 \label{eq:junction-2}
\end{equation}
%
%%%%%%%%%%%%%%%%%%%%%%%%%%%%%%%%%%%%%%%%%%%%%%%%%%%%%%%%%%%%%%%%%%%%%%%%%
which gives a non-local boundary condition for $\Omega$. 
Alternatively, we can combine Eqs.(\ref{eq:effective_Einstein_1}) and
(\ref{eq:effective_Einstein_3}) to get 
%%%%%%%%%%%%%%%%%%%%%%%%%%%%%%%%%%%%%%%%%%%%%%%%%%%%%%%%%%%%%%%%%%%%%%%%%
%
\begin{equation}
  a_0 \kappa_5^2 (\delta \rho - 3H \delta q) 
  =- \frac{k^2}{a_0^2} \left(\Omega'- \frac{a'}{a} \Omega \right)_{\rm b}.
  \label{eq:junction_pre}
\end{equation}
%
%%%%%%%%%%%%%%%%%%%%%%%%%%%%%%%%%%%%%%%%%%%%%%%%%%%%%%%%%%%%%%%%%%%%%%%%%
Then using Eqs.(\ref{eq:scalar_energy_momentum_1}) and
(\ref{eq:scalar_energy_momentum_3}) and rewriting $\delta \phi$
by $Q$, we obtain 
%%%%%%%%%%%%%%%%%%%%%%%%%%%%%%%%%%%%%%%%%%%%%%%%%%%%%%%%%%%%%%%%%%%%%%%%%
%
\begin{equation}
 \kappa_5^2a_0\dot{\phi}^2\left(\frac{H}{\dot{\phi}}Q\right)^{\cdot}
 =-  \frac{k^2}{a_0^2}\left\{
    \frac{\kappa_5^2\dot{\phi}^2}{6}\left(\dot{\Omega}-H\Omega\right)
          + H\left(\Omega'-\frac{a'}{a}\Omega\right)
  \right\}_{\rm b}. 
  \label{eq:junction}
\end{equation}
%
%%%%%%%%%%%%%%%%%%%%%%%%%%%%%%%%%%%%%%%%%%%%%%%%%%%%%%%%%%%%%%%%%%%%%%%%%
In summary, the master equation (\ref{eq:master_equation}), the
Mukhanov-Sasaki equation (\ref{eq:Mukhanov_Sasaki_equation}) with a
source (\ref{eq:Mukhanov_Sasaki_source}) and the junction condition
(\ref{eq:junction-2}) or (\ref{eq:junction}) 
form a closed system for the coupled inflaton and metric perturbations. 

Note that, combining Eqs. (\ref{eq:effective_Einstein_1}) - 
(\ref{eq:effective_Einstein_3}),
one can derive a boundary condition for $\Omega$ which does not include 
$Q$ \cite{Deff}. This boundary condition is equivalent to 
choose (\ref{eq:Mukhanov_Sasaki_equation}) and (\ref{eq:junction-2}) as 
boundary conditions.

%===========================================================================%
%===========================================================================%
\section{Numerical scheme}
\label{sec:scheme}
%===========================================================================%
%===========================================================================%

In this subsection, we briefly discuss the numerical scheme to solve
the equations. First of all, in order to avoid the effect from the
coordinate singularity at $y=y_{\rm c}$ in the GN coordinate, we
introduce an artificial cutoff (regulator) boundary in the bulk at
$y=y_{\rm reg}=\gamma y_{\rm c}$. The constant $\gamma$
controls the position of the regulator brane. 
Since the regulator brane should be far from the physical brane and
$\gamma<1$, we set $\gamma=0.99$ in our numerical simulations.
We discuss the boundary condition at the regulator brane in the 
next section.

In this paper, the numerical calculations of this system were carried
out by employing the pseudo-spectral method \cite{Canuto}. The master
variable $\Omega(t, y)$ is decomposed by the
Tchebychev polynomials defined as
%%%%%%%%%%%%%%%%%%%%%%%%%%%%%%%%%%%%%%%%%%%%%%%%%%%%%%%%%%%%%%%%%%%%%%
%
\begin{equation}
T_n(\xi)\equiv \cos(n\cos^{-1}\xi)\;\;{\rm for}\;\; -1\leq \xi \leq 1,
\end{equation}
%
%%%%%%%%%%%%%%%%%%%%%%%%%%%%%%%%%%%%%%%%%%%%%%%%%%%%%%%%%%%%%%%%%%%%%%
which is a polynomial function of $\xi$ of order $n$. 
Here the variable $\xi$ is related to the GN coordinate $y$. To
implement the pseudo-spectral method, we must transform a spatial
coordinate $y$ in the GN coordinate by
%%%%%%%%%%%%%%%%%%%%%%%%%%%%%%%%%%%%%%%%%%%%%%%%%%%%%%%%%%%%%%%%%%%%%%
%
\begin{equation}
y = \frac{1}{2}\left\{
        \left(y_{\rm reg}-y_{\rm b}\right)\xi +
        \left(y_{\rm reg}+y_{\rm b}\right)
    \right\},
\label{eq:Tchebychev_coordinate}
\end{equation}
%
%%%%%%%%%%%%%%%%%%%%%%%%%%%%%%%%%%%%%%%%%%%%%%%%%%%%%%%%%%%%%%%%%%%%%%
so that the locations of both the physical and the regulator branes
are kept fixed and the spatial coordinate $y$ is projected to the
compact domain $-1\leq \xi \leq 1$. Adopting this coordinate system, the
quantity $\Omega(t,\xi)$ is first transformed into the Tchebychev space
through the relation,
%%%%%%%%%%%%%%%%%%%%%%%%%%%%%%%%%%%%%%%%%%%%%%%%%%%%%%%%%%%%%%%%%%%%%%
%
\begin{equation}
\Omega(t,\xi)= \sum^N_{n=0} \widetilde{\Omega}_n(t)T_n(\xi),
\label{eq:Tchebychev_transform}
\end{equation}
%
%%%%%%%%%%%%%%%%%%%%%%%%%%%%%%%%%%%%%%%%%%%%%%%%%%%%%%%%%%%%%%%%%%%%%%
where we set $N=64$ or $128$. We then discretize the $\xi$-axis to
the $N+1$ points (collocation points)  using the inhomogeneous grid
$\xi_n=\cos(n\pi/N)$ called Gauss-Lobatto collocation
points. With this grid, the fast Fourier transformation can be applied to
perform the transformation between the amplitude $\Omega(t,\xi)$ and the
coefficients $\widetilde{\Omega}_n(t)$. Then the master equation
(\ref{eq:master_equation}) rewritten in the new coordinates are decomposed
into a set of ordinary differential equations for
$\widetilde{\Omega}_n(t)$. For the temporal evolution of
$\widetilde{\Omega}_n(t)$ and $Q(t)$ satisfying 
Eq.(\ref{eq:Mukhanov_Sasaki_equation}), we use Adams-Bashforth-Moulton
method with the predictor-corrector scheme. Further technical details of
the numerical scheme is summarized in the appendix of Ref.~\cite{HT}.

%===========================================================================%
%===========================================================================%
\section{Evolution of curvature perturbations}
\label{sec:numerical}
%===========================================================================%
%===========================================================================%

%-------------------------------------%
%
\subsection{Initial conditions and boundary conditions}
%
%-------------------------------------%
The boundary condition at the brane is given by the junction condition
Eq.~(\ref{eq:junction-2}) or Eq.~(\ref{eq:junction}).
At the regulator brane, we use a Dirichlet boundary condition 
%%%%%%%%%%%%%%%%%%%%%%%%%%%%%%%%%%%%%%%%%%%%%%%%%%%%%%%%%%%%%%%%%%%%%%
%
\begin{equation}
  \Omega|_{y=y_{\rm reg}} = 0. 
  \label{eq:dirichlet}
\end{equation}
%
%%%%%%%%%%%%%%%%%%%%%%%%%%%%%%%%%%%%%%%%%%%%%%%%%%%%%%%%%%%%%%%%%%%%%%
This is because the analysis in de Sitter brane background \cite{KLMW,
KMW} suggests that the inflaton fluctuation excites the bound states
that decays quickly away from the brane. If we focus on this bound state
the result should not be sensitive to the location of the regulator
brane. We will check this fact numerically later. 
Different boundary conditions from Eq.~(\ref{eq:dirichlet}) may produce 
reflection waves from a regulator brane, which affect 
the evolution of Q. A proper boundary condition should be 
imposed on the past Cauchy horizon. This requires a quantisation
of 5D metric perturbations in the bulk, which will turn out to be a highly 
non-trivial problem due to the coupling to the inflaton 
perturbations. 

In 4D cases, $Q$ satisfies 
%%%%%%%%%%%%%%%%%%%%%%%%%%%%%%%%%%%%%%%%%%%%%%%%%%%%%%%%%%%%%%%%%%%%%%%%%
%
\begin{equation}  
  \ddot{Q} + 3H \dot{Q} + \frac{k^2}{a_0^2} Q =0,
  \label{eq:evolution_4D}
\end{equation}
%
%%%%%%%%%%%%%%%%%%%%%%%%%%%%%%%%%%%%%%%%%%%%%%%%%%%%%%%%%%%%%%%%%%%%%%%%%
on small scales since $k^2$ term is dominant over the mass term induced
by the dynamics of the inflaton [see (\ref{eq:Mukhanov_Sasaki_equation})]. 
Thus the solution on small scales is
given by simple plane waves and we can choose the usual 
4D Bunch-Davis vacuum state, which specifies an
appropriate phase and normalization.
On the other hand, in the present 5D case, it is not guaranteed that
the perturbed inflaton field obeys the usual 4D Klein-Gordon equation
due to the coupling to 
the bulk metric perturbations, which is described by $J(\Omega)$
in the right-hand side of Eq. (\ref{eq:Mukhanov_Sasaki_equation}).
In order to obtain the appropriate initial conditions for $Q(t)$, 
we again need a quantisation of the coupled system of the inflaton 
perturbations on a brane and 5D metric perturbations in the bulk.

The aim of this paper is to check whether we can neglect $J(\Omega)$
on small scales or not. If this term could be neglected on the small scales, 
the quantity $a_0 Q$ would behave just as plane waves with a constant
amplitude as in the standard 4D cases.
We call the solution of Eq. (\ref{eq:Mukhanov_Sasaki_equation}) with 
$J(\Omega)=0$ or, equivalently, Eq. (\ref{eq:evolution_4D}) on small scales 
4D solutions. If the solutions for the coupled equations agree with 
4D solutions,
we can assume that the inflaton perturbation behaves
as a free massless field
without taking into account the effect of 5D metric 
perturbations, which is the assumption made in literatures. 

For this purpose, we take the simplest possible initial conditions for
$\Omega(y,t)$ and then specify initial conditions for $Q(t)$ so that they
are consistent with the junction condition (\ref{eq:junction}). We set
$\Omega(y,t_i) = 0$, $\dot{\Omega}(y,t_i) = 0$ and $Q(t_i) = 1$.
Then the time-derivative of $Q(t)$ is determined from the junction
condition (\ref{eq:junction}).

Note that our initial conditions do not give standard qunatum vacuum 
state for 4D solutions. They are merely references to see the effects 
of $J(\Omega)$ on a classical evolution of $Q(t)$. 

%-------------------------------------%
%
\subsection{Evolution of curvature perturbations}
\label{subsec:HL_numerical}
%
%-------------------------------------%

%%%%%%%%%%%%%%%%%%%%%%%%%%%%%%%%%%%%%%%%%%%%%%%%%%%%%%%%%%%%%%%%%%%%%%%%
%%%%%%%%%%%%%%%%%%%%%%%%%%%%%%%%%%%%%%%%%%%%%%%%%%%%%%%%%%%%%%%%%%%%%%%%

We solve numerically the wave equation for the bulk metric perturbations
(\ref{eq:master_equation}) in the GN coordinates with the junction condition 
(\ref{eq:junction}) supplemented by the evolution equation for
inflaton perturbations (\ref{eq:Mukhanov_Sasaki_equation}) .
We set the parameter of the Hawkins-Lidsey model as $C=0.1$ and
the initial time for the simulations as $\mu t_i =40$.
At the initial time, the Hubble scale is given by
%%%%%%%%%%%%%%%%%%%%%%%%%%%%%%%%%%%%%%%%%%%%%%%%%%%%%%%%%%%%%%%%%%%%%%%%%
%
\begin{equation}
 \frac{H(t_i)}{\mu} \approx 2.92.
\end{equation}
%
%%%%%%%%%%%%%%%%%%%%%%%%%%%%%%%%%%%%%%%%%%%%%%%%%%%%%%%%%%%%%%%%%%%%%%%%%
Thus we are considering high-energy region $H/\mu >1$. 

\begin{figure}[ht]
 \centering
 {
   \includegraphics[width=8cm]{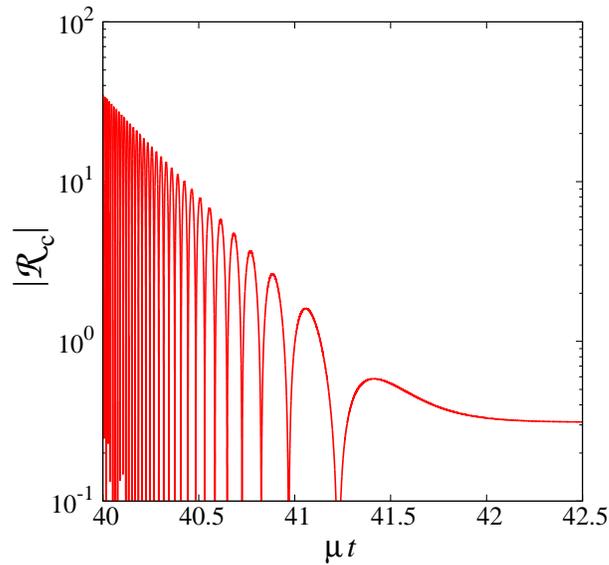}
 }
 \caption{
 The evolution of curvature perturbations 
 ${\cal R}_c$ in the inflationary epoch. We set $C=0.1$, $\mu t_{\rm i}=40$
 and $k/\mu=5$ which corresponds to $k/a_0(t_i)\mu \approx 296$.
 \label{fig:full_R}} 
\end{figure}

\begin{figure}[ht]
 \centering
 {
   \includegraphics[width=14cm]{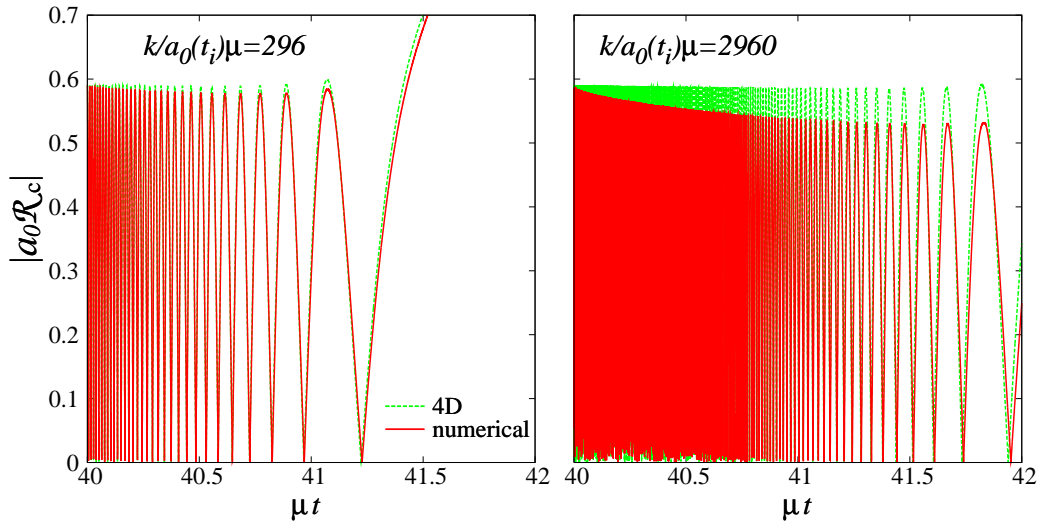}
 }
 \caption{
 The curvature perturbations (multiplied by the scale factor) in the inflationary
 epoch for a long-wavelength mode ({\it left}; $k/a_0(t_i)\mu \approx 296$) 
 and for a short-wavelength mode ({\it right}; $k/a_0(t_i)\mu \approx 2960$). 
 The solid lines represent numerical results and the dashed lines show
 the 4D predictions obtained by neglecting the coupling to the bulk metric 
 perturbations, that is, by solving Eq.~(\ref{eq:Mukhanov_Sasaki_equation})
 with $J(\Omega)=0$. We set $C=0.1$ and $\mu t_i=40$.
 \label{fig:full_aR}} 
\end{figure}

\begin{figure}[ht]
 \centering
 {
   \includegraphics[width=8cm]{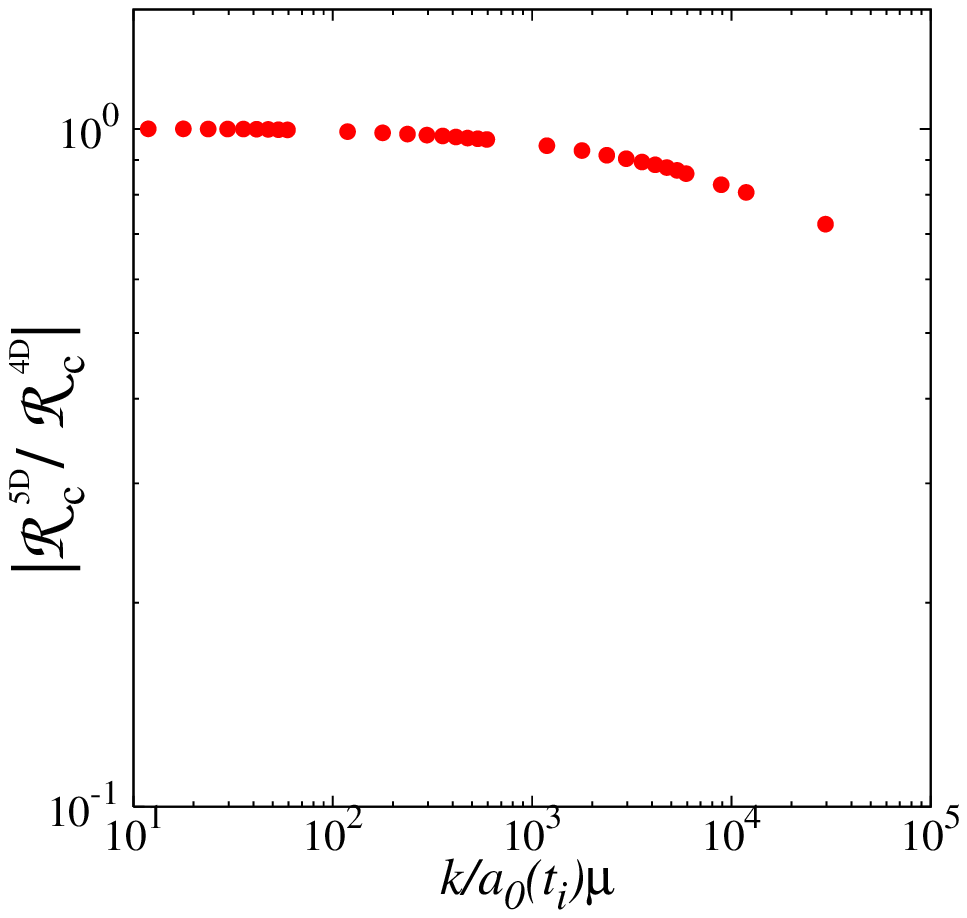}
 }
 \caption{
 The scale dependence of the curvature perturbations evaluated just 
 before the horizon crossing.
 The horizontal axis represents the physical
 scale of perturbations evaluated at the initial time $\mu t_i = 40$.
 We estimated the ratio 
 $|{\cal R}^{\rm 5D}_{\rm c}/{\cal R}^{\rm 4D}_{\rm c}|$ for each wave
 number where ${\cal R}^{\rm 5D}_{\rm c}$ is the curvature perturbation
 in the brane world model and ${\cal R}^{\rm 4D}_{\rm c}$ is the one
 in a 4D model with the identical background dynamics.
  \label{fig:spectrum}} 
\end{figure}

The observable is the comoving curvature perturbation defined by 
%%%%%%%%%%%%%%%%%%%%%%%%%%%%%%%%%%%%%%%%%%%%%%%%%%%%%%%%%%%%%%%%%%%%%%%%%
%
\begin{equation}
  {\cal R}_{\rm c} \equiv -\frac{H}{\dot{\phi}} Q.
\end{equation}
%
%%%%%%%%%%%%%%%%%%%%%%%%%%%%%%%%%%%%%%%%%%%%%%%%%%%%%%%%%%%%%%%%%%%%%%%%%
We focus on the dynamics of this quantity. 
Fig.~\ref{fig:full_R} shows the behaviour of curvature perturbations
for $k/\mu=5$ which corresponds to $k/a_0(t_i)\mu\approx 296$. 
The curvature perturbations ${\cal R}_c$ 
becomes constant on super-horizon scales, which has been shown to be valid 
even in brane world models \cite{cons}. On sub-horizon scales, 
while ${\cal R}_c \propto 1/a_0$ in
the 4D cosmology, a suppression of the amplitude is
observed in the brane-world model, which is due to the coupling to
the bulk metric perturbations. In Fig.~\ref{fig:full_aR}, we
compare the amplitude of ${\cal R}_c$ obtained in numerical 
simulations with the one obtained by neglecting the 
coupling to the bulk metric perturbations, i.e. $J(\Omega)=0$ 
in Eq.~(\ref{eq:Mukhanov_Sasaki_equation}).  
While the difference is very small for 
the long-wavelength modes (left panel; $k/a_0(t_i) \mu=296$), 
the suppression becomes significant for the short-wavelength modes
(right panel; $k/a_0(t_i) \mu =2960$).
Due to this, the spectrum of ${\cal R}_{\rm c}$ just before the horizon 
crossing acquires a scale dependence as is shown in Fig.~\ref{fig:spectrum}. 
Perturbations with larger 
wavenumber stay on sub-horizon scales for a longer time than 
those with smaller wavenumber, so they receive more 
suppression.

\begin{figure}[ht]
 \centering
 {
   \includegraphics[width=7.5cm]{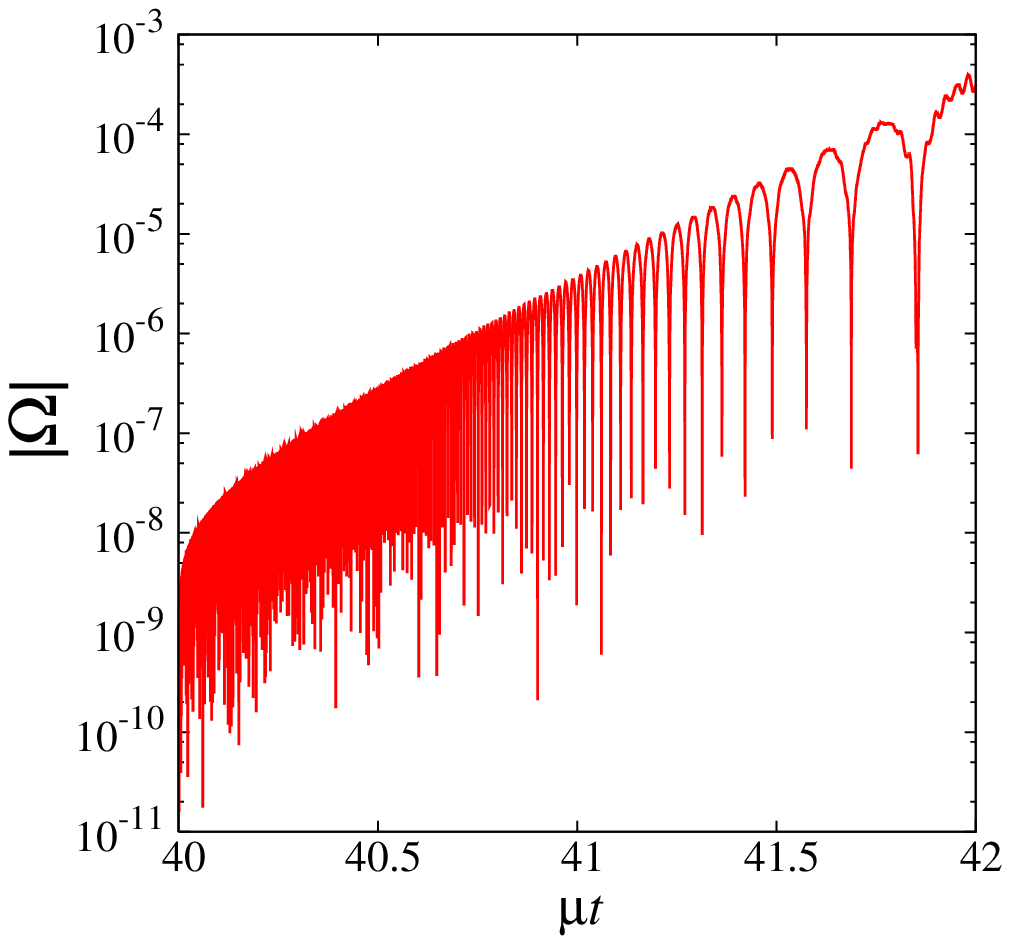}
   \includegraphics[width=7.5cm]{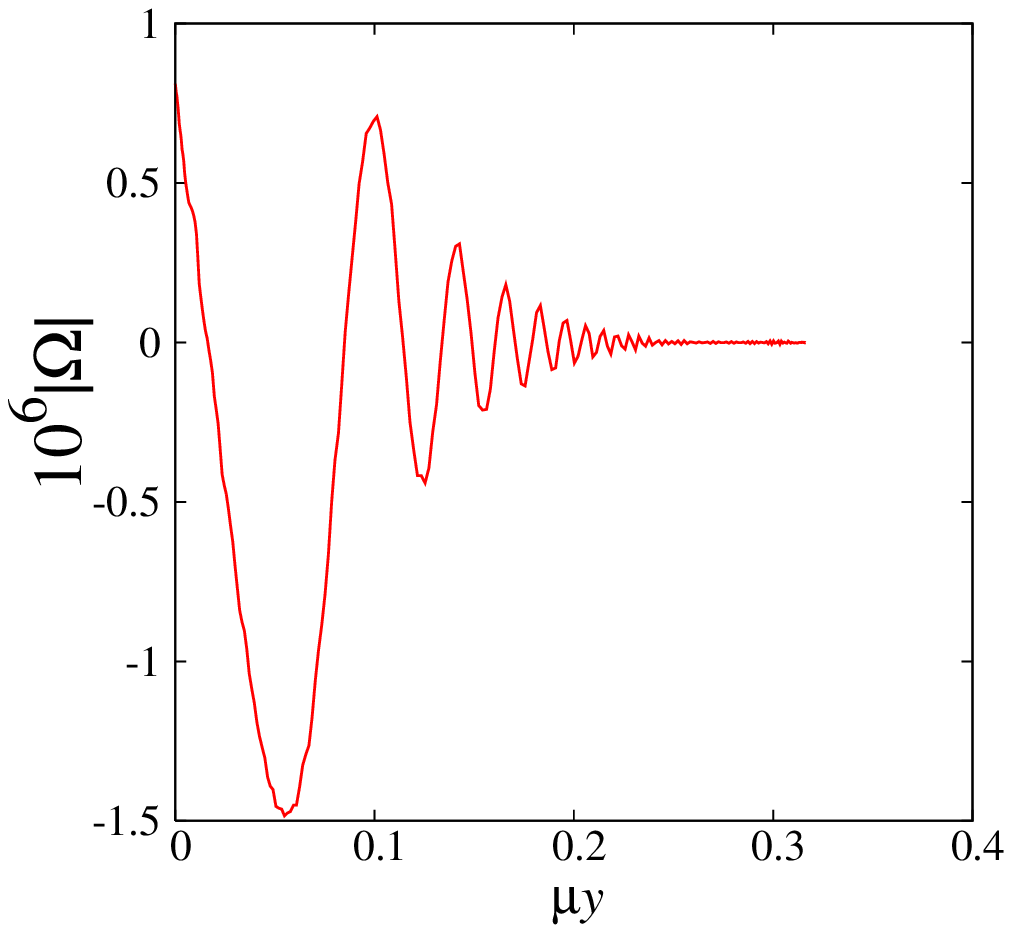}
 }
 \caption{
 {\it Left} : The bulk metric perturbation $\Omega$ evaluated on the
 brane. {\it Right} : The snapshot of $\Omega$ at $\mu t=41.0$. 
 At this time, the coordinate singularity in the bulk is
 located at $\mu y_c(41.0)\approx 0.34$.
 We set $C=0.1$, $\mu t_i=40$, $k/a_0(t_i)\mu=2960$.
 \label{fig:full_Omega}} 
\end{figure}

The suppression of the amplitude  $\mathcal{R}_{\rm c}$ under horizon 
may be understood as the excitation of metric perturbations. The suppression 
of the amplitude of $Q$ is transferred into the enhancement of the metric 
perturbations $\Omega$ in the bulk, which is shown in the left panel of Fig.~\ref{fig:full_Omega}.

Here we comment on the reason why we evaluated $\mathcal{R}_{\rm c}$ just
before the horizon crossing. In the 4D case, we can choose the
Bunch-Davis vacuum as an initial condition. Hence we can determine the
amplitude and phase of the inflaton perturbations and obtain the
spectrum of $\mathcal{R}_{\rm c}$ without any
ambiguities. On the contrary, in the present 5D case, the initial conditions 
cannot be determined without quantising the coupled system. 
Even in the classical level, the junction condition (\ref{eq:junction-2})
or (\ref{eq:junction}) restricts our choice of the initial conditions.
On superhorizon scales, the initial phase of the perturbations essentially
determines its final amplitude after the horizon crossing because the 
phase determines the amplitude of the growing mode solution on 
large scales. Then we get an unphysical k-dependence of the
curvature perturbations evaluated on superhorizon scales 
due to the restricted choice of $dQ/dt$ for given $k$. 
On the other hand, the initial phase is not important when we evaluate
the amplitude of the curvature perturbations on small scales.
Our aim is to see how $J(\Omega)$ term changes the amplitude 
of $Q$ in comparison with the 4D solution for various $k$ on 
small scales. Therefore we chose to calculate 
the spectrum below the horizon scales.

%//////////////////////////////////////////////////////%
%
\subsection{Check of numerical calculations}
\label{subsec:check}
%
%//////////////////////////////////////////////////////%

In this section, we first check the assumption of introducing a regulator.
As expected from the previous work, $\Omega$ decays fast away from the 
brane as is shown in the right panel of Fig.~\ref{fig:full_Omega}. 
This justifies the use of boundary condition 
$\Omega(y_{\rm reg},t)=0$. In addition, thanks to this behaviour of $\Omega$,
the evolution of the curvature perturbations does not depend on the 
location of the regulator. In fact, even at late times when the 
information from the regulator brane comes into the brane, the 
behaviour of the curvature perturbations does not change. Thus 
our result is not obscured by the regulator brane nor the 
existence of the coordinate singularity of the GN coordinate.

Finally, we check the accuracy of our numerical simulations 
by checking the constraint equations derived from the Einstein 
equations. These equations can be generally expressed as
%%%%%%%%%%%%%%%%%%%%%%%%%%%%%%%%%%%%%%%%%%%%%%%%%%%%%%%%%%%%%%%%%%%%%%%%%
%
\begin{equation}
 \mathcal{L}(t)\equiv \sum_k C_k(t)f_k(t) = 0,
 \label{eq:constraint_general}
\end{equation}
%
%%%%%%%%%%%%%%%%%%%%%%%%%%%%%%%%%%%%%%%%%%%%%%%%%%%%%%%%%%%%%%%%%%%%%%%%%
where
%%%%%%%%%%%%%%%%%%%%%%%%%%%%%%%%%%%%%%%%%%%%%%%%%%%%%%%%%%%%%%%%%%%%%%%%%
%
\begin{equation}
  f_k(t) = \{ \ddot{\mathcal{R}}_{\rm b}, \dot{\mathcal{R}}_{\rm b}
 , \mathcal{R}_{\rm b}, \dot{A}_{\rm b}, A_{\rm b}
 , \delta\rho_{\mathcal{E}}, \delta q_{\mathcal{E}}, \delta\pi_{\mathcal{E}}\},
 \label{eq:constraint_quantity_set}
\end{equation}
%
%%%%%%%%%%%%%%%%%%%%%%%%%%%%%%%%%%%%%%%%%%%%%%%%%%%%%%%%%%%%%%%%%%%%%%%%%
and $C_k(t)$ is a set of their coefficients. We calculated
the relative error for the constraint equations (\ref{eq:effective_Einstein_1})
(\ref{eq:effective_Einstein_3}) and (\ref{eq:effective_Einstein_4}) as
%%%%%%%%%%%%%%%%%%%%%%%%%%%%%%%%%%%%%%%%%%%%%%%%%%%%%%%%%%%%%%%%%%%%%%%%%
%
\begin{equation}
 E(t) = \frac{|\mathcal{L}(t)|}{\max_{k}|C_k(t)f_k(t)|}.
 \label{eq:relative_error}
\end{equation}
%
%%%%%%%%%%%%%%%%%%%%%%%%%%%%%%%%%%%%%%%%%%%%%%%%%%%%%%%%%%%%%%%%%%%%%%%%%
Note that the constraint equation (\ref{eq:effective_Einstein_2}) contains
higher derivative terms of $\Omega$ like $\ddot{\Omega}'$, which are not
solved in our simulations. 
%[see, Eq.(\ref{eq:master_equation})(\ref{eq:Mukhanov_Sasaki_equation})
%(\ref{eq:junction-2}) and (\ref{eq:junction})]. 
Hence it is impossible to evaluate the numerical accuracy of 
(\ref{eq:effective_Einstein_2}) from our numerical simulation.
In fact, if we rewrite the higher-derivative terms of $\Omega$ 
by $Q$ using Eq.~(\ref{eq:junction-2}), Eq.~(\ref{eq:effective_Einstein_2}) 
becomes an identity.

\begin{figure}[ht]
 \centering
 {
   \includegraphics[width=7.5cm]{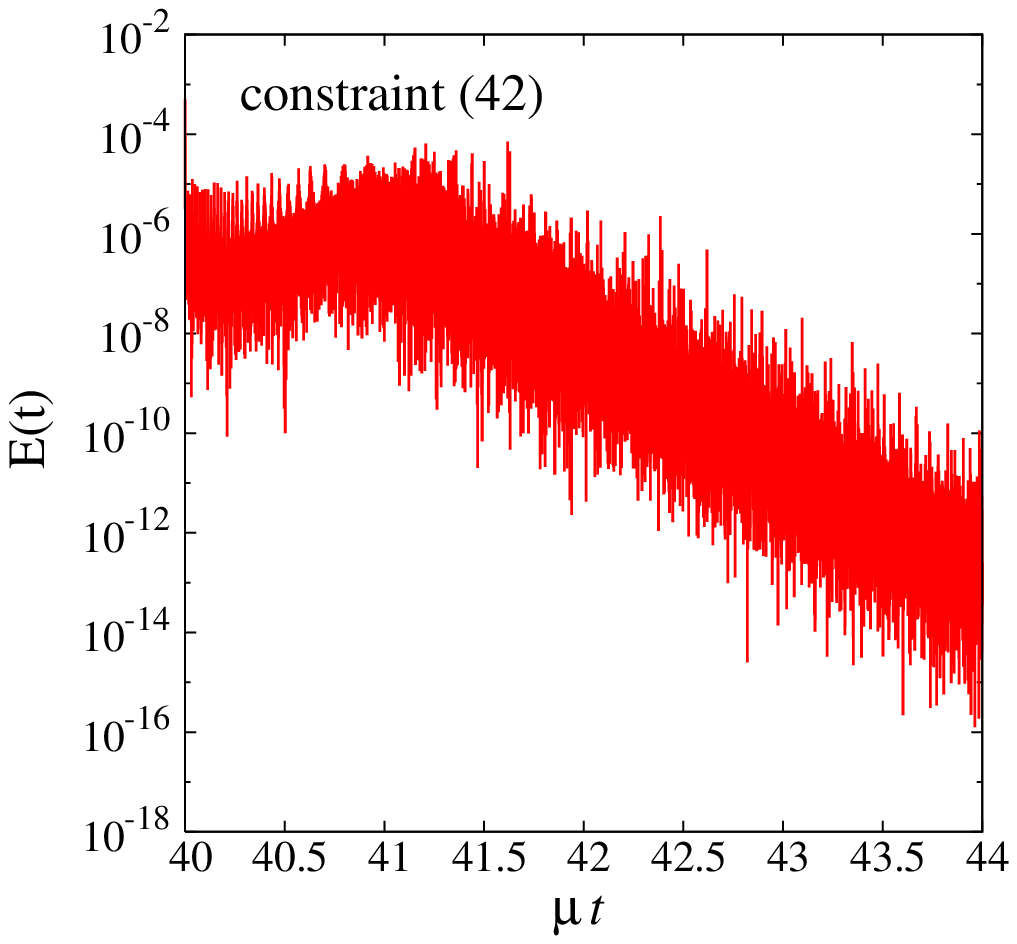}
   \includegraphics[width=7.5cm]{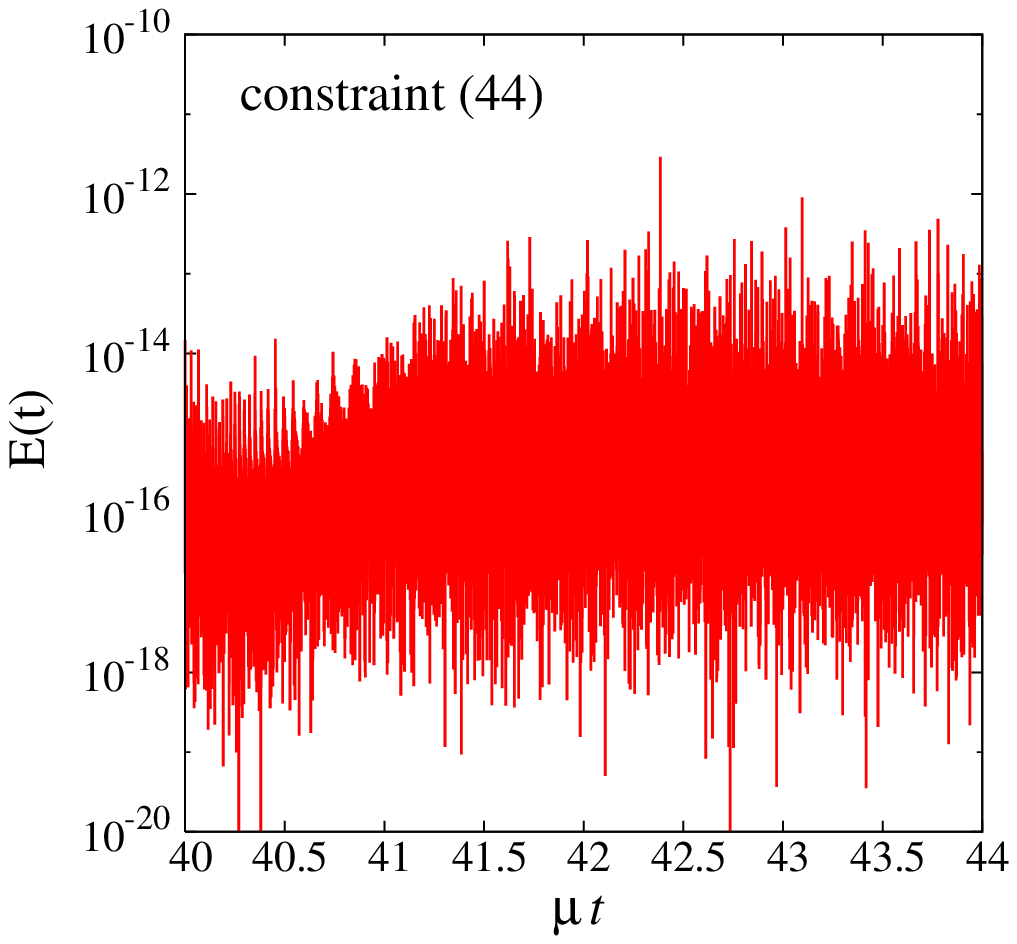}
 }
 \includegraphics[width=7.5cm]{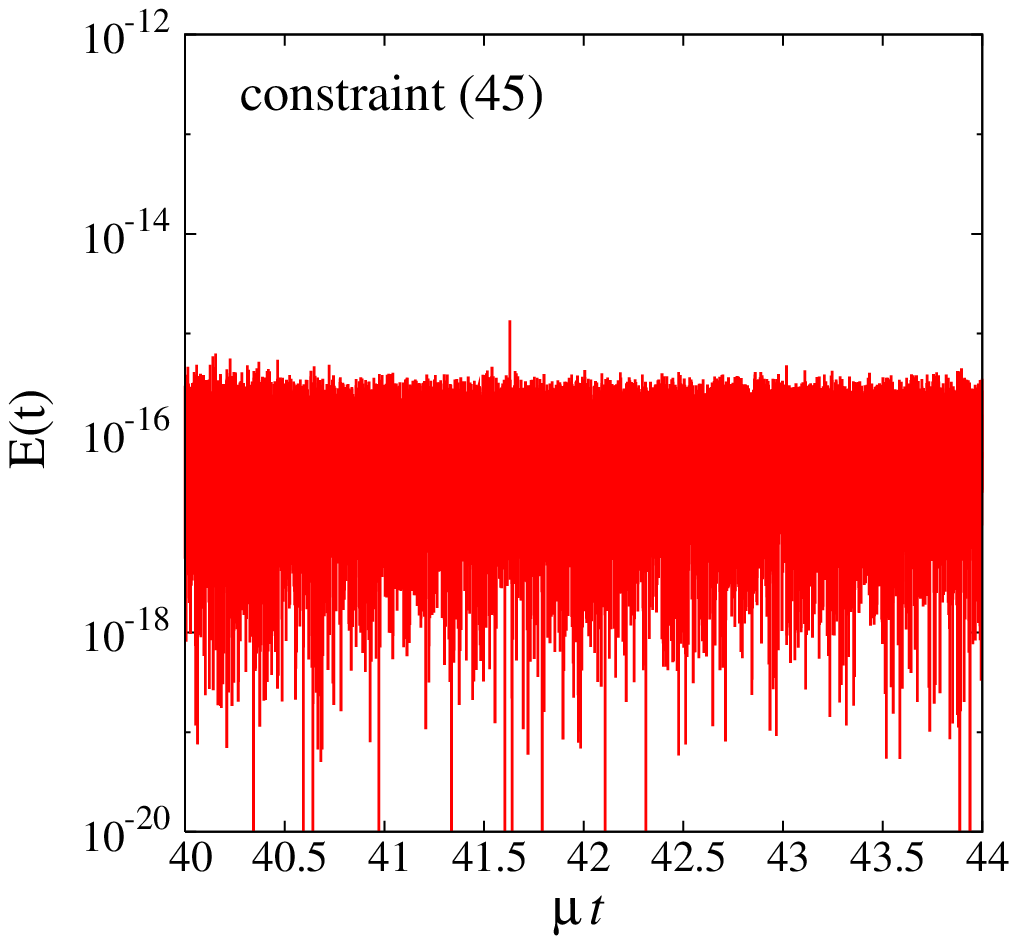}
 \caption{
 The numerical errors of constraint equations
 (\ref{eq:effective_Einstein_1})[{\it upper-left}], 
 (\ref{eq:effective_Einstein_3})[{\it upper-right}] and
 (\ref{eq:effective_Einstein_4})[{\it lower}].
 These plots are results of calculations for $k/a_0(t_i)\mu=296$.
 \label{fig:constraint}} 
\end{figure}

Fig.~\ref{fig:constraint} shows that the relative error of 
Eq. (\ref{eq:effective_Einstein_1}) is suppressed to
less than $10^{-5}$ and Eqs. (\ref{eq:effective_Einstein_3}) and
(\ref{eq:effective_Einstein_4}) are very precisely satisfied, indicating
that our numerical scheme for solving the coupling system
(\ref{eq:master_equation}) and (\ref{eq:Mukhanov_Sasaki_equation}) is
reliable. Note that the constraint equation
(\ref{eq:effective_Einstein_4}) has a
different character from the others, which measures the
numerical accuracy of the master equation (\ref{eq:master_equation})
on the brane.

%===========================================================================%
%===========================================================================%
\section{Conclusion}
\label{sec:concusion}
%===========================================================================%
%===========================================================================%

In this paper, we studied the evolution of the curvature perturbations 
in a brane-world inflation model where an inflaton is living on a single
brane in a 5D AdS spacetime. We used the Hawkins-Lidsey model of 
the inflaton potential which enables us to derive the background 
solutions analytically. We solved the full coupled evolution equations 
for the inflaton perturbations described by the Mukhanov-Sasaki variable
and the bulk metric perturbations described by the master variable. 
This is the first numerical result for the evolution of scalar 
curvature perturbations in a brane-world that consistently takes 
into account the backreaction of the metric perturbations. 

We used the GN coordinate to describe the bulk spacetime that has 
a coordinate singularity. Then we are forced to introduce a regulator 
brane to cut the spacetime. We have checked that the evolution of 
curvature perturbations is insensitive to the location of the 
regulator. This is because the bulk metric perturbations decay 
fast away from the brane. We adopted the simplest possible 
initial condition for $\Omega$ by taking $\Omega =0$ 
and $\dot{\Omega}=0$ at an 
initial time. Then we fix $Q$ so that the boundary condition
is consistent. In order to check whether we can consistently neglect 
the coupling to bulk metric perturbations described by $J(\Omega)$ 
in Mukhanov-Sasaki equation (\ref{eq:Mukhanov_Sasaki_equation}) or not, 
we followed the subsequent evolution of curvature perturbations numerically.  

The evolution of curvature perturbations showed a suppression 
of the amplitude compared with the conventional 4D model
even on sub-horizon scales. This means that it is impossible to 
neglect a coupling to gravity even on small scales due to the coupling 
to the higher-dimensional gravity through $J(\Omega)$. The suppression of 
the amplitude may be understood 
as a loss of energy due to the excitation of the bulk metric 
perturbations as we took the initial condition that
$\Omega(y,t_i)=0$ and $\dot{\Omega}(y,t_i)=0$.
On super-horizon scales, the curvature perturbations
become constant, which confirms the fact that the constancy of 
the curvature perturbations is independent of gravitational theory 
for adiabatic perturbations \cite{cons}. 

Our result suggests that an usual assumption that the inflaton 
perturbations (the Mukhanov-Sasaki variable) approach to a free 
massless field on small scales cannot be applied in a brane world models
on small scales at high energies.
Then the spectrum becomes sensitive to initial conditions 
not only for the inflaton perturbations but also for 
bulk metric perturbations.
In this sense, we should not take the result Fig.~\ref{fig:spectrum}
at face value
because this result is based on a special choice of initial conditions.
In order to determine the initial conditions without ambiguity, 
it is required to quantise 
the coupled brane (inflaton)-bulk (metric perturbations) system 
in a consistent manner along the line of \cite{quantum}. 
This is an outstanding open question to be addressed in order to derive 
the spectrum of the inflaton perturbations
in a brane-world. 

%%%%%%%%%%%%%%%%%%%%%%%%%%%%%%%%%%%%%%%%%%%%%%%%%%%%
\begin{ack}  
TH is supported by JSPS (Japan Society for the Promotion of
Science). KK is supported by PPARC.
\end{ack}    
%%%%%%%%%%%%%%%%%%%%%%%%%%%%%%%%%%%%%%%%%%%%%%%%%%%%

\end{document}